\documentclass[aip,twocolumn,showpacs,amsmath,amssymb,superscriptaddress]{revtex4}

\usepackage{graphicx,dcolumn}
\usepackage{bm}
\usepackage[colorlinks=false]{hyper ref}

\begin{document}

\title{Sample-to-sample fluctuations in real-network ensembles}

\author{Nicole Carlson}  
\affiliation{Department of Physics and Redwood Center for Theoretical Neuroscience, University of California, Berkeley, CA 94720, USA}

\author{Dong-Hee Kim}
\affiliation{Department of Applied Physics, Aalto University, P.O. Box 15100, 00076 AALTO, Finland}

\author{Adilson E. Motter}
\affiliation{Department of Physics and Astronomy and Northwestern Institute on Complex Systems, Northwestern University, Evanston, IL 60208, USA}

\date{\today}

\begin{abstract}
\noindent
Network modeling based on ensemble averages tacitly assumes that the networks meant to be modeled are typical in the ensemble. Previous research on network eigenvalues, which govern a range of dynamical phenomena, has shown that this is indeed the case for uncorrelated networks with minimum degree $\ge 3$.  Here we focus on real networks, which generally have both structural correlations and low-degree nodes.  We show that: (i) the ensemble distribution of the dynamically most important eigenvalues can be not only broad and far apart from the real eigenvalue but also highly structured, often with a multimodal rather than bell-shaped form;  (ii) these interesting properties are found to be due to low-degree nodes, mainly those with degree $< 3$, and network communities, which is a common form of structural correlation found in real networks.   In addition to having implications for ensemble-based approaches, this shows that low-degree nodes  may have a stronger influence on collective dynamics than previously anticipated from the study of computer-generated networks. 

\end{abstract}


\onecolumngrid
\noindent
{\small Published in \href{http://chaos.aip.org/resource/1/chaoeh/v21/i2/p025105_s1}{{\it Chaos} {\bf 21},  025105 (2011)}; DOI:10.1063/1.3602223}\\[-1mm]

\maketitle

\noindent
{\bf 
In the network modeling of collective behavior, while one can analyze each network individually, the ideal is to draw general conclusions that can apply to an entire class of networks.  However, one must ascertain under what conditions such results apply. These can be determined by considering ensembles, and many results have already been established for ensembles of random networks.  It remains to be addressed, though, the extent to which random ensembles are representative of real networks.
Given a real network ${\cal N}$, one can define an associated ensemble ${\cal E_N}(p_1, ...,p_n)$ as  the set of all possible  realizations of the network in which one or more parameters, represented by  $p_1, ...,p_n$, are preserved. In one extreme there is ${\cal E_N}(N)$, where one only fixes the number $N$ of nodes, so that the real network could be very dissimilar from most ensemble elements. In the opposite extreme there is ${\cal E_N}(p_1, p_2, ...)$, where all possible parameters are fixed, but this is equivalent to studying the original network.  An important goal is to restrict as few of the parameters as possible while still capturing the essential features of the real network.  This is fundamental for the study of collective dynamics because in many network processes, including diffusion, consensus phenomena, and synchronization, the  influence of the network structure is determined by the eigenvalues of a coupling matrix, which exhibit a rather convoluted dependence on simple network properties.  Focusing primarily on the ensemble ${\cal E_N}(N, \{k_i\})$, which preserves the number of nodes and the degree sequence $\{k_i\}$, we study the ensemble distribution of individual eigenvalues  and the conditions under which the ensemble networks are representative of the real network.
}

\begin{table*}
\begin{ruledtabular}
{\small
\begin{tabular}{l c c c c c c c c}

{\it  Real Networks} & $N$ & $\langle k \rangle$ & $k_N$
& $\lambda_2^{(0)}$ 
& $\lambda_N^{(0)}$
& $\Lambda_N^{(0)}$
& $\mu_2^{(0)}$ 
& $\mu_N^{(0)}$\\

& ($q_2$, $q_3$) & &
& ($\Delta_{\lambda_2}$, $\sigma_{\lambda_2}$) 
& ($\Delta_{\lambda_N}$, $\sigma_{\lambda_N}$) 
& ($\Delta_{\Lambda_N}$, $\sigma_{\Lambda_N}$) 
& ($\Delta_{\mu_2}$, $\sigma_{\mu_2}$) 
& ($\Delta_{\mu_N}$, $\sigma_{\mu_N}$)\\

\hline

Protein interaction & 4,927 & 6.5 & 296 & 0.117 & 297.02 & 21.64 & 0.058 & 1.93 \\
network (yeast) \cite{Goh2005} & (0.73, 0.53) & & & 
(-1.09, 0.031)$^\dag$ & (-10.83, 0.004) & (-14.49, 0.26) & (-1.73, 0.022)$^\dag$ & (1.30, 0.022)$^\dag$ \\ \hline

Gene regulatory & 662 & 3.2 & 71 & 0.020 & 72.03 & 9.98 & 0.011& 1.99 \\ 
network (\textit{yeast}) \cite{EcoliGeneRegulation} & (0.51, 0.25) & & & 
(-2.75, 0.025) & (-6.34, 0.015) & (-7.34, 0.30) & (-2.71, 0.015) & (2.73, 0.015) \\ \hline

Neural network & 297 & 14.5 & 134 & 0.849 & 135.05 & 24.37 & 0.195 & 1.55 \\ 
(\textit{C.\ elegans}) \cite{Watts1998} & (0.95, 0.94) & & & 
(0.55, 0.100) & (-12.57, 0.002) & (0.50, 0.16) & (-6.07, 0.048) & (0.68, 0.046) \\ \hline

Metabolic network  & 268 & 4.3 & 45 & 0.280 & 46.12 & 8.92 & 0.102 & 1.88 \\ 
(\textit{E.\ coli}) \cite{dataset} & (0.89, 0.45) & & & 
(-0.77, 0.085) & (-0.03, 0.017) & (1.23, 0.21) & (-2.17, 0.022) & (1.32, 0.021)\\ \hline

Food web  & 183 & 26.6 & 105 & 0.980 & 106.15 & 41.31 & 0.207 & 1.82 \\
(Little Rock) \cite{martinez_1991} & (0.99, 0.99) & & & 
(0.39, 0.042)$^*$ & (-1.97, 0.02) & (12.49, 0.12) & (-11.25, 0.035) & (26.84, 0.033)\\ \hline

Coauthorship network & 36,458 & 9.4 & 278 & 0.019  & 279.17 & 51.29 & 0.004 & 1.95 \\
(cond-mat) \cite{Newman2001} & (0.92, 0.78) & & & 
(-4.99, 0.033)$^\dag$  & (0.13, 0.061)$^*$ & (195.43, 0.11) & (-5.10, 0.023)$^\dag$  & (3.09, 0.023)$^\dag$\\ \hline

Political blog & 1,222 & 27.4 & 351 & 0.169 & 352.05 & 74.08 & 0.081 & 1.79 \\
network \cite{Adamic2005}& (0.89, 0.80) & & & 
(-2.22, 0.160)$^*$ & (-27.58, 0.002) & (-3.53, 0.20) & (-3.12, 0.090)$^\dag$ & (1.73, 0.090)$^\dag$\\ \hline

Word network  & 30,243 & 59.9 & 1,145 & 0.111& 1,146.1 & 146.83 & 0.016 & 1.71 \\ 
(Moby Thesaurus) \cite{Motter2002} & (0.99, 0.99) & & & 
(-19.63, 0.041) & (-2.26, 0.005) & (196.06, 0.11) & (-16.17, 0.041) & (9.62, 0.041)\\  \hline

Internet  & 10,515 & 4.1 & 2,277 & 0.091 & 2,278.0 & 58.03 & 0.036 & 1.93 \\
 (autonomous syst.) \cite{dataset} & (0.64, 0.20) & & & 
(1.52, 0.015)$^\dag$ & (0.0, 0.003) & (-9.07, 0.24) & (-0.51, 0.009)$^\dag$ & (2.90, 0.009)$^\dag$\\ \hline

Power grid & 4,941 & 2.7 & 19 & 0.001 & 20.11 & 7.48 & 0.0002 & 1.99 \\
(western U.S.) \cite{Watts1998} & (0.68, 0.02) & & & 
(-3.79, 0.007) & (-1.15, 0.10) & (27.51, 0.10) & (-3.75, 0.004)& (1.66, 0.004)\\ \hline

Electronic circuit & 512 & 3.2 & 22 & 0.029 & 23.11& 5.01 & 0.009 & 1.96 \\
(s838) \cite{dataset}& (0.9, 0.0) & & & (-3.19, 0.033) & (-0.98, 0.021) & (-1.61, 0.08) & (-4.68, 0.012) & (1.75, 0.012)\\ \hline

Airport network & 332 & 12.8	& 139 & 0.120 & 140.03 & 41.23 & 0.035 & 1.72 \\ 
(U.S.) \cite{dataset} & (0.83, 0.68) & & & 
(-2.68, 0.160)$^*$ & (-5.88, 0.005) & (22.04, 0.18) & (-3.85, 0.089)$^\dag$ & (1.13, 0.085)$^\dag$\\ 

\end{tabular}
}
\end{ruledtabular}
\caption{
\label{table1}
Real networks considered in this study. The columns show basic properties of the real networks as well as the extreme eigenvalues,  the corresponding spectral positions $\Delta_x$,  and the standard deviations  $\sigma_x$ of the random ensembles (see Sec.\ \ref{sec3}). 
The basic properties are the number of nodes $N$,  
the average degree $\langle k \rangle$, the maximum degree $k_N$. The minimum degree $k_1$ is one for all  networks.  
In each case, we focus on the largest connected component of the real network. In the $k$-core test with $k=2$ and $k=3$, the percentage of remaining nodes is $q_2$ and $q_3$, respectively.
A summary of the $3$-core tests is given as superscripted symbols,  where $\dag$ indicates an originally structured  
eigenvalue  distribution that becomes unimodal in the $3$-core networks while those with $*$ are still structured
in the $3$-core networks.  No symbol is specified for non-structured distributions in the original random ensemble. 
}
\end{table*}

\section{Introduction}

The complexity of networked systems is frequently studied via  empirical observation that different real networks  share  common structural properties~\cite{netbook,Albert2002}.  Such common properties have implications for network dynamical phenomena, which are often believed not to depend strongly on the specific network under consideration \cite{Albert2002,newman2003}.  Ensembles of networks designed to reproduce common properties, including heterogeneous degree distribution and certain level of randomness, have been widely used in statistical physics studies of networks~\cite{newman2003,Dorogovtsev2008}. This  provides a convenient tool to address general and possibly universal aspects of network phenomena~\cite{Dorogovtsev2008,arenas2008}. The inverse approach, focused on building a precise model to reproduce an observed network dynamical phenomenon, remains challenging in general. But to what extent can ensemble studies provide information about the properties of individual networks?

Previous research focused on the eigenvalues of coupling matrices has shown that the ensemble distribution of  the eigenvalues converge to peaked, bell-shaped functions as the number of nodes in the network is increased \cite{Kim2007}.  This result was established for random uncorrelated networks of given degree distributions with minimum degree $\ge 3$, showing that under these conditions the eigenvalues of large networks are well represented by ensemble averages.  However, the sample-to-sample fluctuations across the ensemble, which determine the quality of ensemble averages,  may change when these conditions are relaxed.  In particular,  it has been shown that having a finite fraction of nodes with degree one or two can fundamentally alter the value of individual eigenvalues in the thermodynamic limit  \cite{Samukhin2008}.
These studies are insightful, and so are those based on the analysis of individual computer-generated networks \cite{Wu2005}. 
Yet, they do not directly address the properties of real networks, since empirically observed networks have finite size,  are structurally correlated, and usually include low-degree nodes.

\begin{figure*}
\includegraphics[width=0.95\textwidth]{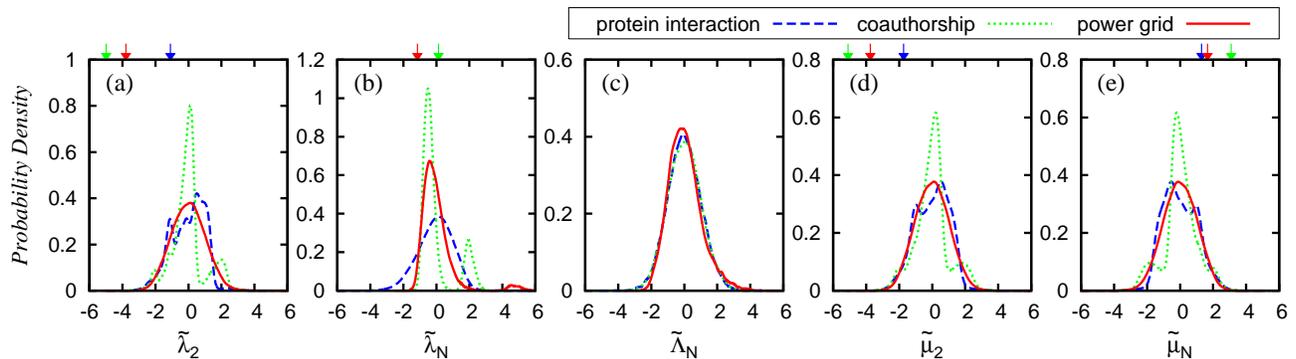}
\caption{
\label{fig1}
Ensemble distributions of extreme eigenvalues for a selection of real networks: 
the protein interaction network,  the coauthorship network, and the 
power grid (Table \ref{table1}). The eigenvalues are 
(a) the smallest nonzero eigenvalue of the Laplacian, 
(b) the largest eigenvalue of the Laplacian, 
(c) the largest eigenvalue of the adjacency matrix, 
(d) the smallest nonzero eigenvalue of the normalized Laplacian, and
(e) the largest eigenvalue of the normalized Laplacian.
Tilde is used to indicate that the distributions are rescaled 
as $\tilde{x} \equiv (x-\langle x \rangle)/\sigma_x$  to have zero averages and unit variances,  
where $\langle x \rangle$ is the average and $\sigma_x$ is the standard deviation of the original distribution $P(x)$
of an eigenvalue $x$. 
The arrows (top) indicate  the positions of the real extreme eigenvalues that lie within the range of the plot, clearly showing
that in most cases the real network is not ``typical" within the random ensemble. 
}
\end{figure*}

The central question that we raise in this context is how typical a real network is in an associated ensemble that preserves a selection of its local structural properties, such as the degree sequence.  Here, we address this question by sampling the associated ensemble and contrasting the relevant eigenvalues of the ensemble elements with those of the real network used to generate it. We focus on the extreme (largest and/or smallest nonzero) eigenvalues of coupling matrices, because they encapsulate the structural network attributes that govern a number of network dynamical processes, such as synchronization~\cite{Barahona2002,Nishikawa2003,Restrepo2006}, diffusion~\cite{Motter2005},  and epidemic spreading~\cite{Boguna2003,Wang2003}.  The results are, therefore, representative of the impact that ensemble-based approaches have on the study of network dynamics in general.

The article is organized as follows. In Sec.\ \ref{sec2}, we introduce and motivate the eigenvalues as well as the real networks we consider. In Sec.\ \ref{sec3}, we show that in some cases the real network is well represented by the ensemble distribution,  but in many other cases the ensemble distribution deviates significantly from the real network.  We also show that the ensemble distributions are often highly structured, exhibiting multiple peaks.  In Sec.\ \ref{sec4}, we explore the properties of $k$-cores and network eigenvectors to elaborate on the origin of these structures. We also discuss the impact of community structures to rationalize the observed deviations of the real eigenvalues from the ensemble distributions. Finally, our concluding remarks are presented in Sec.\ \ref{sec5}.

\section{Eigenvalues and empirical networks}
\label{sec2}

We focus on  the extreme eigenvalues of three connectivity matrices 
that play important roles in many dynamical processes: the adjacency matrix,
the Laplacian matrix, and the normalized Laplacian matrix. 
The adjacency matrix is defined as $A=(A_{ij})$, where $A_{ij}=1$ if nodes $i$ and $j$ 
are connected and $A_{ij}=0$ otherwise. The Laplacian $L$ and 
the normalized Laplacian $\hat{L}$ are defined as $D-A$ and $D^{-1}L$,
respectively, where $D=\mathrm{diag}\{ k_1, \ldots , k_N \}$ is the diagonal matrix of degrees.  
For connected  undirected networks, as  considered here,  the smallest eigenvalue of 
the matrix $L$ is zero and all the others are strictly positive. The same holds true for
the matrix $\hat{L}$. 
These coupling matrices have broad significance for the study of network dynamics.
Synchronization of diffusively coupled
oscillators, for example, is often determined by the largest  eigenvalue ($\lambda_N$) and
the smallest nonzero eigenvalue ($\lambda_2$) of the Laplacian $L$ \cite{Barahona2002,nishpnas}.  The relaxation
time in diffusion processes is determined by the corresponding eigenvalues
of the normalized Laplacian $\hat{L}$ \cite{Motter2005}, which we denote $\mu_N$ and $\mu_2$, respectively. 
The threshold for epidemic spreading \cite{Wang2003} and the dynamic range in excitable systems \cite{rest2011}, 
on the other hand, are  largely influenced by the largest eigenvalue ($\Lambda_N$) 
of the adjacency matrix $A$. Motivated by these and other dynamical applications in 
which  extreme eigenvalues are found to play a role,  the eigenvalues of interest 
in this study are $\lambda_2$, $\lambda_N$, $\mu_2$, $\mu_N$, and $\Lambda_N$. 
For notational convenience, the nodes are labeled in increasing order of their degrees $k_i$,  
such that $k_1 \le \ldots \le k_N$.

We consider twelve real networks from various domains, including technology, biological sciences, and 
sociology \cite{dataset}, which span a wide range of sizes and link densities  (Table \ref{table1}). 
For each of these networks we define the associated {\it random ensemble} ${\cal E_N}(N, \{k_i\})$, 
which preserves the number of nodes and the degree sequence, and we study the properties 
of the extreme eigenvalues in this ensemble.    This is implemented computationally by randomly selecting 
independent network realizations in the ensemble. The ensemble networks are generated using the link-rewiring 
algorithm~\cite{Newman2002}, which randomizes a network while preserving the given
degree sequence $\{k_i\}$. In this construction, all links are regarded as undirected, self-links and duplicated
links are forbidden, and the networks are required to remain connected. Two realizations become statistically
independent  through $(\sum_i k_i)^2$ link rewiring operations.  Our statistics are based
on $10,000$ independent  
network realizations for each ensemble. The finite number of realizations
leads to a discrete set of the extreme eigenvalues $\{ x_i \}$, so that the distribution can be formally written
as $P(x) \propto \sum_i \delta(x-x_i)$.  To avoid artifacts associated with the discreteness of the distribution, 
the  Dirac delta  $\delta(x)$ is approximated  as a Gaussian distribution with a small variance.

\section{Eigenvalue ensemble distributions}
\label{sec3}

\begin{figure}
\includegraphics[width=0.45\textwidth]{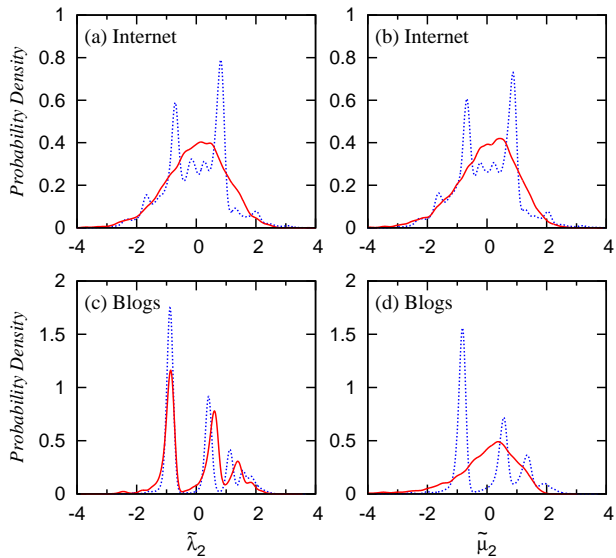}
\caption{
\label{fig2}
Effect of low-degree nodes on  ensemble distributions for the Internet and the network 
of political blogs. The distributions correspond to 
(a, c) the smallest nonzero eigenvalue of the Laplacian
and (b, d) the smallest nonzero eigenvalue of the normalized 
Laplacian. The distribution for the largest eigenvalue of the 
normalized Laplacian is essentially undistinguishable from
the latter, and is not shown.
Dotted lines indicate the distributions 
associated with the original real networks and continuous 
lines indicate the distributions for the corresponding $3$-cores of the networks.
All distributions are rescaled as in Fig.\ \ref{fig1}. In most cases the ensemble 
distributions for the  $3$-cores are significantly smoother than for the original networks, 
indicating that at least part of the observed structures is due to low-degree nodes.
}
\end{figure}

Figure~\ref{fig1} shows the ensemble distributions of the extreme eigenvalues for a selection of 
disparate real networks---a protein-interaction network, a scientific coauthorship network, and 
a power-grid network.  Some of the distributions, such as the largest eigenvalue of the adjacency
matrix for all three networks (Fig.~\ref{fig1}(c)) and the largest normalized Laplacian eigenvalue for the
power-grid network (Fig.~\ref{fig1}(e)), exhibit relatively well-defined bell-shaped distributions.
Others, however, exhibit pronounced deviations, including secondary peaks. This is the case, for 
example, for the extreme eigenvalues of the  Laplacian (Fig.~\ref{fig1}(a)-(b)) and normalized Laplacian (Fig.~\ref{fig1}(d)-(e))
of the coauthorship network.  Additional information is provided by  considering the 
position of the corresponding eigenvalues of the real networks relative to these ensemble distributions.
Surprisingly, in most cases the eigenvalue of the real network deviates significantly from the ensemble
average. A notable exception is the largest  Laplacian eigenvalue of the coauthorship network 
(Fig.~\ref{fig1}(b)), where the real-network is well approximated by the ensemble average. Perhaps 
even more surprisingly, there appears to be essentially no relation between the bell-shaped form of the
distribution and the quality of this approximation. For example, the
largest Laplacian eigenvalue of the protein-interaction network does not lie within the range of the plot
despite the bell-shaped form of the ensemble distribution (Fig.~\ref{fig1}(b)), and the same is true for the 
largest eigenvalue of the adjacency matrix of all three networks (Fig.~\ref{fig1}(c)).
 
To quantify this deviation, we consider the spectral position of the real-network 
eigenvalues, which we define as 
\begin{equation}
\Delta_x = (x^{(0)} - \langle x \rangle ) / \sigma_x, \nonumber
\end{equation}
where $x$ represents the eigenvalue. Here, the superscript $(0)$ indicates the eigenvalue of the
real network, $\langle x \rangle$ is the average of the eigenvalue in the associated 
ensemble, and $\sigma_x$ is the standard deviation of the ensemble distribution, $P(x)$.
This simple quantity provides a meaningful measure for the extent to which a real
eigenvalue deviates from the ensemble average, which is expressed in units of the
standard deviation.

Table \ref{table1} summarizes the statistics for all $12$ empirical networks 
considered. Several properties of 
the real-network eigenvalues, such as the  approximate symmetry between 
$\mu_2$ and  $2-\mu_N$, 
are in good agreement with theoretical predictions \cite{Kim2007}.  
However, in most cases the value of $|\Delta_x|$ is larger than unity, and in
many cases it is much larger, confirming that real networks are often
not typical in their own associated ensembles. That is, in terms of the extreme 
eigenvalues considered here, the real networks are often significantly different from the
majority of the ensemble networks. 
Another interesting aspect of the results 
shown in Table \ref{table1} is that this deviation is not necessarily due to large 
deviations in absolute values.  For all real networks,  $\lambda_N$ is just
slightly larger than $k_N+1$, as predicted theoretically  for uncorrelated 
networks  \cite{Kim2007}. The ensemble distributions are also peaked close to
this point (at a distance $<10^{-2}\times\lambda_N^{(0)}$ for all networks),
but because these distributions are very narrow, a small deviation
in absolute value tends to correspond to a relatively large number in units of 
standard deviation.

On the other hand, several cases exhibit a structured rather  than bell-shaped distribution, which
cannot be anticipated from these theoretical results.   
This is so for the smallest nonzero eigenvalues of the Laplacian
and of the normalized Laplacian  (the same is true also for the largest eigenvalue 
of the normalized Laplacian, due to symmetry mentioned above, which is nearly
exact for the ensemble networks). 
These cases are marked with superscripted  symbols  in Table \ref{table1}. 
For example, for the Internet network and the network of political blogs,  the ensemble 
distributions of these eigenvalues  exhibit multiple large and relatively distant peaks (Fig.\ \ref{fig2}).  
The next question concerns
the origin of these  abnormal fluctuations.  We hypothesize that the main cause of the 
fluctuations is the presence of poorly connected nodes  and/or poorly connected  groups 
of nodes. The basis for this hypothesis
is that the smallest nonzero eigenvalues of Laplacian-like matrices are known to be
influenced by low-degree nodes \cite{Samukhin2008} as well as by communities of densely
connected nodes that are sparsely connected with the rest of the network \cite{newman2008}. 
Next, we study the extent to which these factors can generate the observed fluctuations
in the ensemble distributions and deviations between  the real eigenvalues and the ensemble averages.

\section{Role of low degrees and additional network structure}
\label{sec4}

To probe the influence of low-degree nodes, we explore the $k$-core organization of the networks \cite{Dorogovtsev2006}.  The  $k$-core of a network is the largest connected subnetwork in which all nodes have degree at least $k$.  Given a real network ${\cal N}$, we extract the $k$-core(${\cal N}$)  and then generate another random ensemble ${\cal E}_{k\mbox{-core}({\cal N})} (N', \{k'_i\})$, where  $N'$ is the number of nodes and $\{k'_i\}$ is the degree sequence of the $k$-core. The first case of interest corresponds to  $2$-cores, where the minimum degree in the new network is $2$. For the networks considered in this study, the $2$-cores are found to exhibit spiky ensemble distributions comparable to those of the original networks. 
Our analysis of $3$-cores, on the other hand, reveals very different behavior: as illustrated in Fig.\ \ref{fig2}(a)-(b) for the Internet network, the distributions of the smallest nonzero eigenvalues of the Laplacian and normalized Laplacian become significantly smoother and close to bell-shaped curves. Similar smoothening of ensemble distributions for the  $3$-cores is observed for  the smallest nonzero eigenvalues of most networks with fluctuations, as summarized in Table~\ref{table1}.  This confirms that the fluctuations in $\lambda_2$ and  $\mu_2$ (and also $\mu_N$)  are mainly due to nodes with degree $1$ and $2$. 

We do not systematically consider  $k$-cores for higher $k$ because the loss of statistics due to the reduction in the size of the network  may compete with the effect of removing low-degree nodes. However, there are cases where the fluctuations still appear in the ensemble distributions of the  $3$-cores, such as for the smallest nonzero eigenvalue of the Laplacian in the network of political blogs (Fig.~\ref{fig2}(c)). This suggests that other types of network structures are affecting some of the ensemble distributions. 

\begin{figure}
\includegraphics[width=0.45\textwidth]{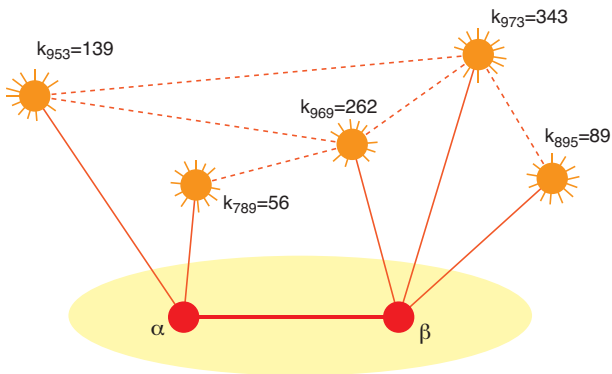}
\caption{Example of network structure contributing to the fluctuations in the distribution $P(\lambda_2)$ associated with the network of political blogs. The subgraph shown    highlights the relevant part of an ensemble network at the second peak (left to right) of the $3$-core random ensemble in  Fig.\ \ref{fig2}(c).  The positions of the peaks in the distribution can be estimated by considering only the submatrix of the Laplacian that includes the  links between the two low-degree nodes ($\alpha$ and $\beta$) and their neighbors (solid lines).  
\label{fig3}
}
\end{figure}

In the particular case of the network of political blogs, we find that there is a relationship between the distribution of the eigenvalue $\lambda_2$  and subgraph structures involving low-degree nodes. Specifically, at the peaks of the distribution of $\lambda_2$, the components of the associated eigenvector are dominantly large for a certain pair of low-degree nodes that are directly connected and whose other neighbors have considerably larger degrees.  Figure \ref{fig3} highlights this structure in an ensemble element that is at one of the peaks of  $P(\lambda_2)$. While different ensemble realization will have different such nodes connected to each other, we can show that the impact they have on the fluctuations of the eigenvalue distributions is mainly determined by their degrees. Overall, there exist only few links between low-degree nodes in a chosen network realization, but the frequency with which an ensemble network exhibits at least one such subgraph is relatively high. This is likely related to the fact that the network of political blogs has a very long-tailed degree distribution, which along with the constraints of not having self-links and duplicated links,  leads to the relatively frequent occurrence of such subgraphs in the ensemble. Moreover, the observation of these subgraphs allows us to estimate analytically the positions of the  peaks of $P(\lambda_2)$ for this network.

Assuming that only one such subgraph contributes to the eigenvector of $\lambda_2$, we can project the full Laplacian onto the reduced space that consists of two low-degree nodes, $\alpha$ and $\beta$, and their neighbors. 
Accordingly, by writing the eigenvalue equation $(L- \lambda_{2} I )\vec{y} = 0$ explicitly and noting that the degrees of the  neighbors of $\alpha$ and $\beta$ are much larger than $\lambda_{2}$, we derive the approximate expression 
\begin{equation}
\lambda_{2} \approx \frac{(k_\alpha + k_\beta)}{2} - \sqrt{\left(\frac{k_\alpha-k_\beta}{2}\right)^2 + 1} - \epsilon_{\alpha\beta}, \nonumber
\label{eq:lambda}
\end{equation}
where $\epsilon_{\alpha\beta} = f_{\alpha\beta}\sum^\prime k_i^{-1}$ is a small number, with $f_{\alpha\beta} = 1+(k_\alpha+k_\beta)/\sqrt{(k_\alpha-k_\beta)^2+4}$ and the summation taken over the neighbors of $\alpha$ and $\beta$. Even with the rough approximation $\epsilon_{\alpha\beta} = 0$, the calculated $\lambda_{2}$ shows remarkable agreement with the peaks of the eigenvalue distribution observed in the network of political blogs. In the random ensemble of the original network, the low-degree combinations provide $\lambda_{2} \approx 0.38$, $0.58$, $0.69$ for $k_\alpha = 1$ and $k_\beta=2$, $3$, $4$, respectively, which are in precise agreement with the observed major peaks of $P(\lambda_2)$. Even though $\epsilon_{\alpha\beta}$ increases with $k_\alpha$ and $k_\beta$,  the equation above also provides very good estimations for the peaks in the ensemble of $3$-cores. The estimations for $k_\alpha = 3$ and $k_\beta=3$, $4$, $5$ are $2.00$, $2.38$, $2.58$, respectively, which are very close to the major peaks observed at $\lambda_2 = 1.96$, $2.33$, $2.53$.  While this eigenvector analysis applies to the network of political blogs, the multimodal distributions found in the other network ensembles may be determined by other network structures. But we suggest that even in such cases, the peaks in the eigenvalue distributions are likely to be associated with patterns of subgraph structures that can take a relatively small number of forms.

\begin{figure}
\medskip
\includegraphics[width=0.45\textwidth]{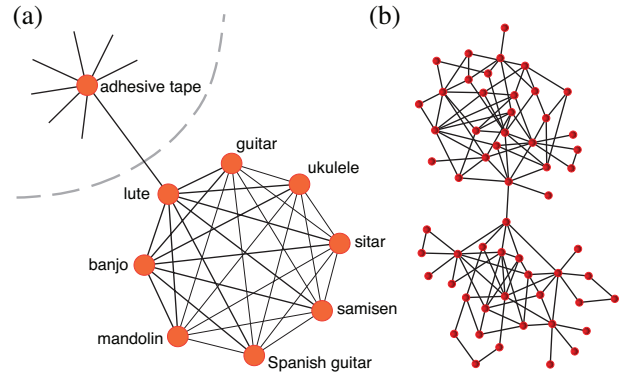}
\caption{
\label{fig4}
Network structure effecting the extreme eigenvalues $\lambda_2$ 
and $\mu_2$. (a) Community structure found in the word network. 
(b) Model network with $50$ nodes, consisting of two clusters 
connected to each other by a single link. The randomization of 
the whole network without preserving the clusters tends to increase
the smallest nonzero eigenvalues of the Laplacian and normalized Laplacian.
}
\end{figure}

Another important question concern the origin of the often large deviation of the real eigenvalues from the ensemble averages even when the ensemble distributions are approximately bell-shaped. We propose that this is caused by the presence of structures in the real networks that would correspond to rare events in the random ensemble. An example of a particularly important such structure that can lead to large deviations from the ensemble averages is shown in Fig.\ \ref{fig4}(a): a densely connected community in the word network  (Moby Thesaurus). This cluster dominantly contributes to the eigenvectors corresponding to the extreme eigenvalues $\lambda_2$ and $\mu_2$.  In the cluster, eight words closely related to `guitar' form a fully connected subnetwork; this cluster is connected to  the rest of the network by the link between `lute', which has two very distinct meanings, and `adhesive tape'. This type of weakly connected network structure can cause  the smallest nonzero eigenvalue of the Laplacian to be very small. This explains the small values of $\lambda_2$ and $\mu_2$ found in the word network. 

To exemplify this effect, consider the model network with two communities shown in  Fig.~\ref{fig4}(b).  The  eigenvalue $\lambda_2$ for this network is $0.045$. For the associated random ensemble, the ensemble average of this eigenvalue is $0.36$, which is substantially larger than the eigenvalue of the initial network. This is so mainly because the ensemble does not preserve the community structure. Indeed, a  community-preserving ensemble can be created in which the two communities are separately randomized and then linked together by a single link,  and within this ensemble the average of $\lambda_2$ is very close  to the eigenvalue of the original network (difference $<10^{-3}$). 

Because this type of structure is expected to be present to some extent in most real networks, the smallest nonzero eigenvalues of the Laplacian and normalized Laplacian of real networks are generally expected to be smaller  than the corresponding random-ensemble averages. This explains the negative spectral position in most cases shown in Table  \ref{table1}. This, in turn, is consistent with the positive spectral position exhibited by the largest eigenvalue of the normalized Laplacian for the networks we consider. Community structures are expected to also impact other eigenvalues, such as the eigenvalues of the adjacency matrix, which have been used to design algorithms for community detection \cite{chauhan_per_2009}. The deviation of the eigenvalues from the ensemble distributions can also be partially determined by different network structures that set them apart from random, such as clustering, degree correlations, and assortative mixing \cite{restrepo_pre_2007,Mieghem}.  Disassortative networks, for example, are known to exhibit enhanced synchronization properties precisely because they have smaller ratio $\lambda_N/\lambda_2$  than their random counterparts \cite{motter_epl_2005,sorrentino_2007}.

\section{Final remarks}
\label{sec5} 

The fluctuations in the ensemble distributions observed in this study have important 
implications. On one hand,  we have provided evidence that these structures are 
largely due to low-degree nodes in the network. On the other hand, it follows that
these fluctuations cannot be ignored in the estimation and interpretation of ensemble
averages associated with networks that have low-degree nodes, which is the rule 
and not the exception among real networks. 
Moreover, because these distributions can be broad, if one samples networks from 
the ensemble, the eigenvalue fluctuations from sample to sample will frequently be large.
Another interesting aspect of this problem
is that low degrees alone may not explain all the observed fluctuations and that, even 
for bell-shaped distributions the real eigenvalue of interest often deviates significantly
from the ensemble average. For some of the eigenvalues, this deviation can be mainly
attributed to the presence of community structures in the network. This, in turn,
suggests that a promising approach would be to incorporate the community
structures in the definition of the ensemble, as additional properties $\{p_i\}$ in 
${\cal E_N}(N, \{k_i\}, \{p_i\})$. There are in fact models to generate random network
ensembles with a large number of preserved properties, including communities 
\cite{bianoni} and distributions of subgraphs \cite{newman}.  An important 
challenge for future research is to address the properties of real network with the
framework provided by such models.

\section*{Acknowledgments}

The authors thank Jie Sun for providing feedback on the manuscript. 
This study was supported by the U.S.\ National Science Foundation under Grants No. DMS-0709212 and No. DMS-1057128, 
and the Academy of Finland under Grant No.\ 139514.

\end{document}